\documentclass[dvipdfm]{elsart}

\usepackage{graphicx, epsfig}
\usepackage{amsmath,amssymb}
\usepackage{rotating}
\usepackage[latin1]{inputenc}

\newcommand {\tr} {\mbox{tr}}    
\def\R{\bibitem{}}

\begin{document}

\begin{frontmatter}
\title{Improved testing inference in mixed linear models}
\author{Tatiane F.~N.~Melo, Silvia L.~P.~Ferrari}
\address{Departamento de Estatística, Universidade de São Paulo, Rua do Matão, 1010,
São Paulo/SP, 05508-090, Brazil}
\author{Francisco Cribari-Neto}
\address{Departamento de Estatística, Universidade Federal de Pernambuco,
Cidade Universitária, Recife/PE, 50740-540, Brazil}
\begin{abstract}
Mixed linear models are commonly used in repeated measures studies. 
They account for the dependence amongst observations obtained from the same 
experimental unit. Oftentimes, the number of observations is small, and it is 
thus important to use inference strategies that incorporate small sample 
corrections. In this paper, we develop modified versions of the likelihood 
ratio test for fixed effects inference in mixed linear models. In particular,
we derive a Bartlett correction to such a test and also to a test obtained from
a modified profile likelihood function. Our results generalize those in Zucker
et al.~({\it Journal of the Royal Statistical Society B\/}, 2000, {\bf 62},
827--838) by allowing the parameter of interest to be vector-valued.
Additionally, our Bartlett corrections allow for random effects nonlinear
covariance matrix structure. We report simulation results which show that
the proposed tests display superior finite sample behavior relative to the
standard likelihood ratio test. An application is also presented and discussed.   

\begin{keyword}
Bartlett correction, Fixed effects, Likelihood ratio test, 
Mixed linear models, Modified profile likelihood function.
\end{keyword}
\end{abstract}
\end{frontmatter}

\section{Introduction}\label{S:intro}

In recent years repeated measures data have been widely analyzed in many 
fields, including biology and medicine. In such studies, the observations are obtained 
from different experimental units, each unit being observed more 
than once (Brown and Prescott,~2006). In particular, some of these studies 
use longitudinal data (Verbeke and Molenberghs, 2000),
in which the observations are collected over time. Mixed linear models have been 
extensively used by practitioners to analyse repeated measures since they 
account for within units correlation.
It is also noteworthy that there is available software specifically designed 
for the estimation of such models; see Pinheiro and Bates~(2000) and
Littel et al.~(2006).

A common shortcoming lies in the fact that in many studies the sample size 
is small which renders approximate inferential procedures 
unreliable. Improved inference may be based on the theory of higher order
asymptotics. Practical applications of such theory may be found in
Brazzale et al.~(2007).
The likelihood ratio test, which is commonly used
to make infe\-ren\-ce on the fixed effects parameters, quite often
displays large size distortions when the sample size is small.
This happens because its null 
distribution is poorly approximated by the limiting $\chi^2$ distribution, 
from which critical values are obtained. It is possible to obtain a Bartlett 
correction factor and use it to modify the likelihood ratio test statistic in such
a way to bring its null distribution closer to its limiting counterpart; the 
approximation error is reduced from $O(n^{-1})$ to $O(n^{-2})$, where $n$ 
is the sample size, thus making any size distortion vanish at a faster rate. 

Another shortcoming relates to the effect of the nuisance parameters on the 
resulting inference on the parameters of interest. Different modifications 
to the profile likelihood function have been proposed with the aim of reducing such 
effect. For a review see Severini~(2000, Chapter 9); see also Sartori et al.~(1999) and
Sartori~(2003).
The adjustment proposed by Cox and Reid~(1987) can be used whe\-ne\-ver 
the nuisance and interest parameters are orthogonal. 
DiCiccio and Stern~(1994) have shown that the Cox--Reid test statistic can 
be Bartlett-corrected, just as the likelihood ratio test statistic.
The combined use of modified profile likelihoods and Bartlett correction can 
deliver accurate and reliable inference in small samples, as evidenced by 
the results in Ferrari et al.~(2004), Ferrari et al.~(2005) and Cysneiros and
Ferrari~(2006).

Zucker et al.~(2000) obtained improved likelihood ratio testing inference 
by deriving Bartlett corrections to the profile and modified (Cox--Reid) 
profile likelihood ratio tests on the fixed effects parameters
in mixed linear models.
Their results, however, are only applicable for testing
one parameter at a time, since they only allow for a scalar parameter of interest.
In many studies, nonetheless, practitioners wish to perform joint testing inference
on a set of parameters, especially when comparing three or more treatments in medical trials.
Also, they derived the Bartlett correction to the profile likelihood ratio test
only for the situation where the covariance matrix for the random effects
has a linear structure. Hence, their results are not fully applicable in many situations
of interest, e.g. when the responses of a single subject are measured sequentially and the
errors are assumed to be autocorrelated.
Our chief goal is to generalize their results so that they are 
valid in situations where the parameter of interest is vector-valued and
the covariance matrix for the random effects is allowed to have a non-linear structure. We 
obtain the Cox--Reid profile likelihood adjustment, and also Bartlett correction 
factors for the profile and adjusted profile likelihood ratio test statistics.

The paper unfolds as follows. Section \ref{S:mixed} introduces the mixed 
linear model, Section \ref{S:corrections} contains the three improved tests 
(Cox--Reid and Bartlett-corrected tests), and Section \ref{S:montecarlo} presents 
simulation study on the finite sample behavior of the standard likelihood 
ratio test and its modified counterparts. An application that uses real 
data is presented and discussed in Section \ref{S:application}. Finally, 
Section \ref{S:conclusions} concludes the paper. Technical details are collected 
in two appendices.  

\section{Mixed linear models}\label{S:mixed} 

The mixed linear model is given by 
  \begin{equation}\label{E.2.0}
{\bf y}_i = X_i{\bf\beta} + Z_i {\bf b}_i + {\bf\epsilon}_i, \: \: \: i = 1,\ldots, N, 
\end{equation}
where ${\bf y}_i = (y_{i1}, y_{i2}, \ldots, y_{i\tau_{i}})^\top$ is a 
$\tau_i \times 1$ vector of responses on the $i$th experimental unit, 
${\bf\beta}$ is an $n$-vector of fixed effects parameters, $X_i$ is a $\tau_i \times
n$ known matrix, ${\bf b}_i$ is a random effects vector ($q \times 1$),
$Z_i$ is a known $\tau_i \times q$ matrix,  
and ${\bf\epsilon}_i = (\epsilon_{i1}, \epsilon_{i2}, \ldots, \epsilon_{i\tau_{i}})^\top$ 
is a $\tau_i \times 1$ vector of random errors. It is oftentimes assumed that 
$\epsilon_i\sim{\mathcal{N}}_{\tau_i}({\bf 0},\sigma^2 I_{\tau_i})$, where $I_{\tau_i}$ 
denotes the $\tau_i \times \tau_i$ identity matrix and ${\bf 0}$ is a vector 
of zeros. It is also assumed that   
${\bf b}_i \sim{\mathcal N}_{q}({\bf 0},G)$, where ${\bf b}_1, {\bf b}_2,
\ldots, {\bf b}_N, {\bf\epsilon}_1, {\bf\epsilon}_2, \ldots, {\bf\epsilon}_N$ are 
independent and $G = G({\bf \varrho})$ is a $q \times q$ positive definite matrix, 
${\bf \varrho}$ being an $m\times 1$ vector of unknown parameters. 
Model (\ref{E.2.0}) can be written in matrix form as 
\begin{equation}\label{E.2.1}
{\bf Y} = X{\bf\beta} + Z{\bf b} + {\bf \epsilon}, 
\end{equation} 
where ${\bf Y} = ({\bf y}^\top_1, {\bf y}^\top_2, \ldots, {\bf y}^\top_N)^\top$
is $T \times 1$, with $T = \sum_{i=1}^{N} \tau_i$, $X = ({\bf X}^\top_1,
{\bf X}^\top_2, \ldots, \break {\bf X}^\top_N)^\top$ is a $T \times n$ matrix, $Z$ is
a $T\times Nq$ diagonal matrix given by $Z =
{\rm diag}(Z_1, Z_2, \ldots, Z_N)$, ${\bf b} = ({\bf b}^\top_1,
{\bf b}^\top_2, \ldots, {\bf b}^\top_N)^\top$ is an $Nq$-vector and
${\bf \epsilon} = ({\bf \epsilon}^\top_1, {\bf \epsilon}^\top_2, \break\ldots,
{\bf \epsilon}^\top_N)^\top$ is $T \times 1$. Thus,
${\bf b}\sim{\mathcal N}_{Nq}({\bf 0},I_N\otimes G)$, where $\otimes$ denotes the 
Kronecker product and ${\bf \epsilon}\sim{\mathcal N}_{T}({\bf 0}, \sigma^2 I_{T})$; 
${\bf b}$ and ${\bf \epsilon}$ are independent.

It is possible to write model (\ref{E.2.1}) as 
\begin{equation}\label{E.2.2}
{\bf Y} = X{\bf\beta} + {\bf e}, 
\end{equation} 
where ${\bf e} = Z {\bf b} + {\bf\epsilon}$. Hence, ${\bf e}\sim{\mathcal N}_{T}({\bf 0},
{\bf \Sigma})$, where ${\bf \Sigma} = {\bf \Sigma}({\bf \omega}) = Z(I_N\otimes G)Z^\top +
\sigma^2 I_{T}$, ${\bf \omega} = ({\bf \varrho}^\top,\sigma^2)^\top$ being an
$(m+1) \times 1$ vector of unknown parameters. 
Hence, the log-likelihood function for model (\ref{E.2.2}) can be expressed as  
\begin{equation}\label{E.2.3}
{\ell}({\bf\beta}, {\bf\omega};{\bf Y}) = -\frac{T}{2}\log(2\pi) - \frac{1}{2}\log
|{\bf \Sigma}| - \frac{1}{2}({\bf Y} - X{\bf\beta})^\top{\bf \Sigma^{-1}}
({\bf Y} - X{\bf\beta}),  
\end{equation}
where $|\cdot |$ denotes matrix determinant. 

Let $\theta = ({\bf\psi}^\top, {\bf\varsigma}^\top, {\bf\omega}^\top)^\top$ be 
the $(n+m+1)$-vector of parameters, where ${\bf\psi} = (\beta_1, \beta_2,
\ldots, \beta_p)^\top$ is the $p$-vector ($p\le n$) containing the first $p$ 
elements of ${\bf\beta}$ and $({\bf\varsigma}^\top, {\bf\omega}^\top)^\top$ 
is the $(n-p+m+1)\times 1$ vector of nuisance parameters with  
${\bf\varsigma} = (\beta_{p+1}, \beta_{p+2}, \ldots, \beta_n)^\top$. 
In what follows we shall focus on fixed effects inference. In particular, 
we wish to test ${\mathcal H}_0: {\bf\psi} = {\bf\psi}^{(0)}$ against
${\mathcal H}_1: {\bf\psi} \neq {\bf\psi}^{(0)},$
where ${\bf\psi}^{(0)}$ is a given $p$-vector.

We follow Zucker et al.~(2000) and use a reparameterization in which the nuisance 
($({\bf\varsigma}^\top, {\bf\omega}^\top)^\top$) and interest (${\bf\psi}$)
parameters are orthogonal. In particular, we transform 
$\theta = ({\bf\psi}^\top, {\bf\varsigma}^\top, {\bf\omega}^\top)^\top$ into
$\vartheta =({\bf\psi}^\top, {\bf\xi}^\top, {\bf\omega}^\top)^\top$, with 
\begin{equation}\label{E.2.4}
 {\bf\xi} = {\bf\varsigma} + (\widetilde{X}_{n-p}^\top {\bf \Sigma^{-1}} \widetilde{X}_{n-p})^{-1}
 \widetilde{X}_{n-p}^\top{\bf \Sigma^{-1}}\widetilde{X}_p {\bf\psi}, 
\end{equation}
where $\widetilde{X}_p$ denotes the 
matrix formed out of the first $p$ columns of 
$X$ and $\widetilde{X}_{n-p}$ contains the remaining $(n-p)$ columns of $X$. It is 
easy to show that 
${\bf\psi}$ is orthogonal to ${\bf\phi}=({\bf\xi}^\top, {\bf\omega}^\top)^\top$, i.e., 
the expected values of 
$\partial^2 {\ell}(\vartheta;{\bf Y})/\partial{\bf\psi}
\partial{\bf\xi}^\top$ and $\partial^2 {\ell}(\vartheta;{\bf Y})/\partial{\bf\psi} \partial{\omega_j}$,
for $j = 1, 2, \ldots, m+1$, are matrices of zeros. By partitioning $X$ as 
$(\widetilde{X}_p, \widetilde{X}_{n-p})$ and ${\bf\beta}$ as
$({\bf\psi}^\top, {\bf\varsigma}^\top)^\top$, we can write $X{\bf\beta} =
\widetilde{X}_p {\bf\psi} + \widetilde{X}_{n-p}{\bf\varsigma}$. Using (\ref{E.2.4})
we obtain 
\begin{equation}
X{\bf\beta} = \widetilde{X}_p' {\bf\psi} + \widetilde{X}_{n-p}{\bf\xi}, 
\nonumber
\end{equation}
where $\widetilde{X}_p' = [I_T - \widetilde{X}_{n-p}(\widetilde{X}_{n-p}^\top{\bf \Sigma^{-1}}\widetilde{X}_{n-p})^{-1}
\widetilde{X}_{n-p}^\top{\bf \Sigma^{-1}}]\widetilde{X}_p$.
It follows that the log-likelihood function in 
(\ref{E.2.3}) can be written as 
\begin{equation}\label{E.2.5.1}
{\ell} = {\ell}(\vartheta;{\bf Y}) = -\frac{T}{2}\log(2\pi) - \frac{1}{2}\log
|{\bf \Sigma}| - \frac{1}{2}{\bf z}^\top {\bf \Sigma^{-1}}{\bf z},
\end{equation}
where ${\bf z} = {\bf z}({\bf Y}, X, \vartheta) = {\bf Y} - \widetilde{X}_p' {\bf\psi} - \widetilde{X}_{n-p}{\bf\xi}$.

\section{Improved likelihood ratio tests}\label{S:corrections}

\subsection{Bartlett correction}\label{S:corrections1}

The profile likelihood function, which only involves the vector of parameters 
of interest, is defined as 
${\ell}_p({\bf \psi}) = {\ell}({\bf \psi}, {\bf \widehat{\phi}}
({\bf \psi}))$, where ${\bf \widehat{\phi}}({\bf \psi})$ is the maximum likelihood 
estimator of 
${\bf \phi}$ for a fixed value of ${\bf \psi}$. The likelihood ratio statistic for testing 
${\mathcal H}_0$ is  
 \begin{equation*}\label{E.2.5.2}
LR = LR({\bf\psi}^{(0)}) = 2\, \left\{{\ell}_p({\bf \widehat{\psi}}) - {\ell}_p({\bf \psi}^{(0)})\right\},
\end{equation*}
where ${\bf \widehat{\psi}}$ denotes the maximum likelihood estimator of ${\bf \psi}$. 
Under the standard regularity conditions and under 
${\mathcal H}_0$, $LR$ converges in distribution to $\chi^2_{p}$.
This first order approximation may not work well in small samples, however. 
In order to achieve more accuracy, Bartlett (1937) proposed multiplying 
$LR$ by a constant, $(1+C/p)^{-1}$, thus obtaining what is now known as 
the Bartlett-corrected test statistic: 
\begin{equation*}\label{E.2.5.3}
LR^* = \frac{LR}{1+C/p},
\end{equation*}
where $C$ is a constant of order $n^{-1}$ chosen such that, under ${\mathcal H}_0$,
${\rm E}(LR^*) = p + O(n^{-3/2})$. In regular problems, and under the 
null hypothesis $LR^*$ is $\chi^2_{p}$ distributed up to an error of order 
$n^{-2}$; see Barndorff-Nielsen and Hall (1988). A general expression for 
$C$ in terms of log-likelihood cumulants up to the fourth order was obtained 
by Lawley~(1956).

One of our goals is to obtain the Bartlett correction term $C$ for testing 
${\mathcal H}_0: \psi=\psi^{(0)}$ against ${\mathcal H}_1: \psi \ne \psi^{(0)}$
for mixed linear models. This is done in Appendix A using Lawley's results;
see (A.1). For simplicity, here we only give the expression
for $C$ when the $\psi^{(0)}={\bf 0}$, which is common in practical applications:  
\begin{equation}\label{Eq.9}
C = \tr\left(D^{-1}\left(-\frac{1}{2} M + \frac{1}{4} P
- \frac{1}{2} \left(\gamma + \nu\right)\tau^\top\right)\right),
\end{equation}
where $\tr(\cdot)$ is the trace operator. Here, $D$, $M$ and $P$
are $(m+1)\times(m+1)$ matrices given by
$$D = \{(1/2)\:\tr({\bf \dot{\Sigma}}^j {\bf \dot{\Sigma}}_{k})\},$$
$$M = \{\tr((\widetilde{X}_p'^\top {\bf \Sigma^{-1}} \widetilde{X}_p')^{-1}
(\widetilde{X}_{p}'^\top {\bf \ddot{\Sigma}}^{jk}\widetilde{X}_{p}' +
2\dot{{X}}_k'^\top {\bf \dot{\Sigma}}^{j}\widetilde{X}_{p}'))\},$$
$$P = \{\tr((\widetilde{X}_{p}'^\top {\bf \dot{\Sigma}}^{j}\widetilde{X}_{p}')
(\widetilde{X}_p'^\top {\bf \Sigma^{-1}} \widetilde{X}_p')^{-1}
(\widetilde{X}_{p}'^\top {\bf \dot{\Sigma}}^{k}\widetilde{X}_{p}')
(\widetilde{X}_p'^\top {\bf \Sigma^{-1}} \widetilde{X}_p')^{-1})\}$$ 
and $\tau$, $\gamma$ and $\nu$ are $(m+1)$-vectors whose $j$th elements are 
$\tr((\widetilde{X}_p'^\top {\bf \Sigma^{-1}} \widetilde{X}_p')^{-1}
\break(\widetilde{X}_{p}'^\top {\bf  \dot{\Sigma}}^{j}\widetilde{X}_{p}'))$, 
$\tr(D^{-1} A^{(j)})$ 
and 
$\tr((\widetilde{X}_{n-p}^\top {\bf \Sigma}^{-1} \widetilde{X}_{n-p})^{-1}
(\widetilde{X}_{n-p}^\top {\bf \dot{\Sigma}}^{j} \widetilde{X}_{n-p}))$,
respectively.  Note that we give the $(j,k)$ element of each matrix.
In our notation, $A^{(j)}$ is the $(m+1)\times(m+1)$ matrix
given by
\begin{equation*}
A^{(j)} =  \{ (1/2)\: \tr({\bf \dot{\Sigma}}^l {\bf \ddot{\Sigma}}_{jk})
- (1/2)\: \tr({\bf \dot{\Sigma}}^k {\bf \ddot{\Sigma}}_{jl}) - (1/2)\: \tr({\bf \dot{\Sigma}}^j
{\bf \ddot{\Sigma}}_{lk}) \}.
\end{equation*}
Also,
${\bf \dot{\Sigma}}_j
= {\partial{\bf \Sigma}}/{\partial{\bf\omega}_j}$, ${\bf \dot{\Sigma}}^j =
{\partial{\bf \Sigma}^{-1}}/{\partial{\bf\omega}_j} = - {\bf \Sigma}^{-1} 
{\bf \dot{\Sigma}}_j {\bf \Sigma}^{-1}$,
${\bf \ddot{\Sigma}}_{jk} =  {\partial^2{\bf \Sigma}}/
{\partial{\bf\omega}_j\partial{\bf\omega}_k}$,
${\bf \ddot{\Sigma}}^{jk} = {\partial^2{\bf \Sigma}^{-1}}/{\partial{\bf\omega}_j\partial{\bf\omega}_k}
= -2 {\bf \dot{\Sigma}}^k {\bf \dot{\Sigma}}_j {\bf \Sigma^{-1}} 
- {\bf \Sigma^{-1}} {\bf \ddot{\Sigma}}_{jk} {\bf \Sigma^{-1}}$
and
$\dot{{X}_j'}={\partial \widetilde{X}_p'}/{\partial{\bf\omega}_j} 
= - \widetilde{X}_{n-p}(\widetilde{X}_{n-p}^\top{\bf \Sigma^{-1}}\widetilde{X}_{n-p})^{-1} 
\widetilde{X}_{n-p}^\top{\bf \dot{\Sigma}}^j
\widetilde{X}_p'.$ 
 
It is noteworthy that (\ref{Eq.9}) generalizes 
the result in Zucker et al.~(2000, eq.\ (3)). Their expression is only valid when  
the parameter under test is scalar and the covariance matrix for the
random effects has a linear structure and so does ${\bf \Sigma}$, i.e.,
${\bf \Sigma}= \sum{\bf\omega}_j Q_j$, where $Q_j$ are known matrices. 
Note that when ${\bf \Sigma}$ has a linear structure we have
${\bf \dot{\Sigma}}_{j} = Q_j, \forall j,$ 
${\bf \ddot{\Sigma}}_{jk} = 0, \forall j, k$, and equation $(\ref{Eq.9})$ becomes  
\begin{equation}\label{Eq.9.1}
C = \tr\left(D^{-1}\left(-\frac{1}{2} M + \frac{1}{4} P
- \frac{1}{2} \nu\tau^\top\right)\right).
\end{equation}
Additionally, when $\psi$ is scalar, our expression $(\ref{Eq.9.1})$ reduces 
to equation (3) in Zucker et al.~(2000).
Also, when the null hypothesis is
${\mathcal H}_0: {\bf\beta}={\bf\beta}^{(0)}$, $(\ref{Eq.9.1})$ reduces to
\begin{equation*}
C = \tr\left(D^{-1}\left(-\frac{1}{2} M_1 + \frac{1}{4} P_1\right)\right),
\end{equation*}
where $$M_1 = \{\tr((X^\top {\bf \Sigma^{-1}} X)^{-1}(X^\top 
{\bf \ddot{\Sigma}}^{jk}X)) \}$$
and
$$P_1 = \{\tr((X^\top {\bf \dot{\Sigma}}^{j}X)(X^\top 
{\bf \Sigma^{-1}} X)^{-1}(X^\top {\bf \dot{\Sigma}}^{k}X)(X^\top 
{\bf \Sigma^{-1}} X)^{-1})\}.$$

\subsection{Cox--Reid profile likelihood adjustment}\label{S:corrections2}

Cox and Reid~(1987) proposed an adjustment to the profile likelihood function 
which can be used when the nuisance and interest parameters are orthogonal. 
The Cox--Reid adjusted profile log-likelihood function is given by 
\begin{equation*}
{\ell}_{pa}({\bf \psi}) = {\ell}_p({\bf \psi}) -
\frac{1}{2} \log \left\{\left|-{\ell}_{\bf{\phi\phi}}\left(\widehat{\phi}({\bf \psi})
\right)\right|\right\},
\end{equation*}
where ${\ell}_{\bf{\phi\phi}}$ is the matrix of second derivatives of 
${\ell}$ with respect to ${\bf \phi}$. The corresponding likelihood ratio 
test statistic is 
\begin{equation*}\label{E.2.10}
LR_{CR}({\bf\psi}^{(0)}) = 2\:\left\{{\ell}_{pa}({\bf \widetilde{\psi}}) - {\ell}_{pa}({\bf \psi}^{(0)})\right\},
\end{equation*}
where $\widetilde{\psi}$ is the maximizer of ${\ell}_{pa}({{\psi}})$. 

The Cox--Reid test statistic is $\chi^2_{p}$ distributed under ${\mathcal H}_0$ up
to an error of order $n^{-1}$, just like the standard likelihood ratio test statistic. 
DiCiccio and Stern~(1994) defined a Bartlett correction to this test statistic 
which reduces the order of the approximation error to $O(n^{-2})$. The 
corrected test statistic is 
\begin{equation*}
LR^*_{CR} = \frac{LR_{CR}}{1+C^*/p},
\end{equation*}
where $C^*$ is a constant of order $n^{-1}$ such that, under ${\mathcal H}_0$,
${\rm E}(LR^*_{CR}) = p + O(n^{-3/2})$. A general expression for 
$C^*$ can be found in DiCiccio and Stern~(1994, eq.\ (25)). In Appendix B, we 
obtain $C^*$ for testing ${\mathcal H}_0$ in mixed linear models; see 
(B.1). Here, we give the expression for $C^*$ for the case where
${\bf \psi}^{(0)} = {\bf 0}$: 
\begin{equation}\label{E.2.12}
C^* = \tr\left(D^{-1}\left\{- M + \frac{1}{4} P +  \gamma^*\tau^\top\right\}\right),
\end{equation}
where $D$, $M$, $P$ and $\tau$ were given above and 
the $j$th element of the vector $\gamma^*$ is $\tr(D^{-1} C^{(j)})$, with
$C^{(j)}$ being an $(m+1)\times(m+1)$ matrix given by
\begin{equation*}
C^{(j)} = \{-\tr({\bf \dot{\Sigma}}^k {\bf \dot{\Sigma}}_j {\bf \Sigma^{-1}}
{\bf \dot{\Sigma}}_l) + (1/2) \; \tr({\bf \dot{\Sigma}}^j {\bf \ddot{\Sigma}}_{kl})
+ (1/2) \; \tr({\bf \dot{\Sigma}}^k {\bf \ddot{\Sigma}}_{jl}) \}.
\end{equation*}
Our expression for $C^*$ generalizes the result in Zucker et al.~(2000, eq.\ (4)),
since their formula is only valid when $p=1$. We notice that
their formula remains valid when the covariance matrix for the random effects
has a nonlinear structure. As expected, $(\ref{E.2.12})$ reduces to
equation (4) of Zucker et al.~(2000) when $p=1$. Also, for testing
${\mathcal H}_0: {\bf\beta}={\bf\beta}^{(0)}$ against 
${\mathcal H}_1: {\bf\beta} \ne {\bf\beta}^{(0)}$, $C^*$
reduces to (\ref{E.2.12}) with $M$ and $P$ replaced by $M_1$ and $P_1$,
respectively, and $\tau$ replaced by $\tau_1$, with
$\tau_1$ being the $(m+1)$-vector whose $j$th element is 
$\tr((X^\top {\bf \Sigma^{-1}} X)^{-1}(X^\top {\bf  \dot{\Sigma}}^{j}X))$.

The expressions we give for $C$ and $C^*$ in (\ref{Eq.9}) and (\ref{E.2.12}), 
respectively, only involve simple operations on vectors and matrices. Therefore, 
they can be easily computed with the aid of a programming language or software
which can perform such operations, e.g. {\tt Ox} (Cribari-Neto and Zarkos, 
2003; Doornik, 2006) and {\tt R}
(Ihaka and Gentleman, 1996). We note that $C$ and $C^*$ only depend on $X$, on the
inverse covariance matrix ${\bf \Sigma}^{-1}$, on the covariance 
matrix ${\bf \Sigma}$ and its first two derivatives with respect to 
$\omega$. 

\section{Simulation study}\label{S:montecarlo}

In this section we shall present the results of Monte Carlo simulation 
experiments in which 
we evaluate the finite sample performances of the likelihood ratio test ($LR$), 
its Bartlett-corrected version ($LR^*$), the adjusted profile likelihood ratio 
test ($LR_{CR}$) and its Bartlett-corrected counterpart ($LR^*_{CR}$). 

The simulations were based on the following mixed linear model: 
\begin{equation*}\label{Modelo.Simulacao}
{\bf y}_{ij} = \beta_0 + \beta_1 t_{ij} + \beta_2 x_{1i} + \beta_3 x_{2i} + b_{0i} + b_{1i} t_{ij}
+ {\bf\epsilon}_{ij},
\end{equation*}
for $j = 1, 2, \ldots, \tau_i$ with $\tau_i \in \{2, 3, 4, 5, 6, 7, 8, 9\}$ and 
$i = 1, 2, \ldots, N$.
The values of $t_{ij}$ were obtained as random draws from the standard uniform distribution 
${\mathcal U}(0,1)$; $x_{1i}$ and $x_{2i}$ are dummy variables. The 
fixed effects parameters are 
$\beta_0, \beta_1, \beta_2, \beta_3$. Also, ${\bf b}_i = (b_{0i} \:\:
b_{1i})^\top\sim{\mathcal N}_{2}({\bf 0},G)$ with 
\begin{equation}\label{MatrizG}
G = \left[\begin{array}{cc} \omega_1 & \omega_2  \\ \omega_2 & \omega_3  \\ \end{array}\right].
\end{equation}
Additionally, the $\epsilon_{ij}$'s are independent from the $b_i$'s, and 
$\epsilon_{i}\sim{\mathcal N}_{\tau_i}(0,\omega_4 I_{\tau_i})$. 
We test 
${\mathcal H}_0: \psi = {\bf 0}$ against 
${\mathcal H}_1: \psi \neq {\bf 0}$, where $\psi = (\beta_2 \:\: \beta_3)^\top$.
   
All simulations were performed using the \texttt{Ox} matrix programming language
(Cribari-Neto and Zarkos, 2003; Doornik, 2006). The number of Monte Carlo replications was 5,000 and the 
sample sizes considered were $N = 12, 24$ and $36$. The parameter values are 
$\beta_0 = 0$, $\beta_1 = 0.2$, $\beta_2 = 0$, $\beta_3 = 0$, $\omega_1 = 1$, 
$\omega_2 = 0$ and $0.25$, $\omega_3 = 0.5$ and $1$, and $\omega_4 = 0.05$. 
All tests were carried out at the following nominal levels: 
$\alpha = 5\%$ and $\alpha = 10\%$. 

The null rejection rates of the four tests under evaluation are displayed in 
Table 1. We note that the likelihood ratio test is liberal. 
For instance, when $\omega_2 = 0$, $\omega_3 = 
0.50$, $N = 12$ and $\alpha = 10\%$, its rejection rate exceeds $20\%$.
It is noteworthy that the three alternative tests outperform the standard 
likelihood ratio test. For $N = 12$ and $N = 24$, the two best performing 
tests are $LR_{CR}$ and $LR^*_{CR}$; $LR^*$ is slightly oversized. For example, 
when $\omega_2 = 0$, $\omega_3 = 0.50$, $N = 12$ and $\alpha = 5\%$, the 
null rejection rates of $LR_{CR}$, $LR^*_{CR}$ and $LR^*$ are, respectively, 
$4.5\%$, $5.3\%$ and $7.6\%$ ($LR$: $13.0\%$). It is not possible to single 
out a global winner between $LR_{CR}$ and $LR^*_{CR}$. When $N = 36$,
the Cox--Reid and the two Bartlett-corrected tests still outperform $LR$; here, 
$LR^*$ slightly outperforms the other two alternative tests, $LR^*_{CR}$ 
being the second best performing test.

\begin{table}[htp]
\renewcommand{\arraystretch}{1.1}                       
\begin{center}
\caption{\small Null rejection rates of the tests of 
${\mathcal H}_0: \psi = {\bf 0}$; entries are percentages.}\label{T:rejections1}
\medskip
{\small
\begin{tabular}{cccccccccccc}\hline 
&&&\multicolumn{4}{c}{$\alpha = 5\%$}&&\multicolumn{4}{c}{$\alpha = 10\%$}
\\\cline{4-7}\cline{9-12}
{$N$}&{$\omega_2$}&$\omega_3$&$LR$&$LR^*$&$LR_{CR}$&$LR^*_{CR}$&&$LR$&$LR^*$&$LR_{CR}$&$LR^*_{CR}$\\\hline
12   &  0 &0.50 &13.0&7.6&4.5&5.3&&20.8&13.1& 9.2&10.2  \\ 
     &  0 &  1  &13.4&7.8&4.8&5.9&&21.7&13.5& 9.6&10.8  \\  
     &0.25&0.50 &11.2&6.0&3.4&4.1&&19.0&11.2& 7.5&8.5   \\
     &0.25&  1  &13.8&7.9&5.1&5.8&&21.9&13.9& 9.6&10.7  \\ \hline 
24   &  0 &0.50 & 8.3&5.6&4.7&5.0&&14.6&10.9& 9.5&10.0  \\ 
     &  0 &  1  & 8.5&5.8&4.9&5.1&&14.6&11.1&10.1&10.5  \\ 
     &0.25&0.50 & 8.6&5.7&4.8&5.1&&14.8&11.1& 9.6&10.2  \\
     &0.25&  1  & 8.7&6.0&4.8&5.1&&15.0&11.4&10.1&10.6  \\ \hline  
36   &  0 &0.50 & 6.4&4.6&4.2&4.4&&12.8&10.1& 9.5&9.8   \\ 
     &  0 &  1  & 6.1&4.9&4.4&4.7&&12.6&9.8 & 9.0&9.4   \\
     &0.25&0.50 & 6.7&4.8&4.3&4.6&&12.4&10.0& 9.3&9.6   \\
     &0.25&  1  & 6.4&4.7&4.3&4.4&&12.6& 9.8& 9.1&9.4   \\ 
\hline
\end{tabular}
}
\end{center} 
\end{table}

\begin{figure}[h]
\centering  
\includegraphics[width=10cm, height=8cm]{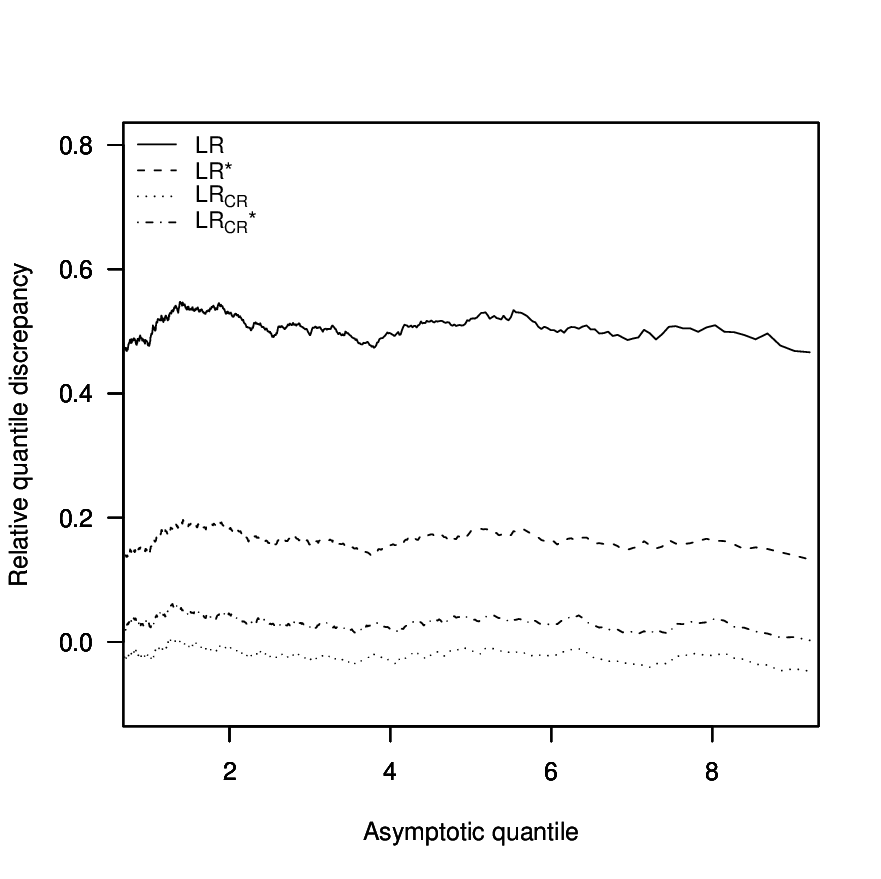}
\caption{\footnotesize Relative quantile discrepancies plot: $N = 12$, $\omega_2 = 0$ 
and $\omega_3 = 0.50$.}\label{fig:quantil1}
\end{figure}

Figure \ref{fig:quantil1} plots the relative quantile discrepancies against the 
asymptotic quantiles for $N=12$, the smallest sample size, where the corrections 
are mostly needed. Relative quantile discrepancies are defined as 
differences between exact
and asymptotic ($\chi^2_2$) 
quantiles divided by the latter. The closer to zero these discrepancies, the 
better the approximation used in the test. We note that the test statistics 
with the smallest relative quantile discrepancies are $LR_{CR}$ and $LR^*_{CR}$. 
We also note that quantiles of $LR$ are approximately 50\% larger than the 
respective asymptotic ($\chi^2_2$) quantiles. 

\vspace{-0.2cm}

Note that the simulated model and the hypothesis under test have practical 
applications, for instance, when the practitioner wishes to compare two different 
treatments and the experimental units are observed in different points in time. 
Here, we assume that the time horizon of the study is limited. This is 
why we used a bounded distribution for choosing values for $t_{ij}$. 
We performed simulations under other situations. We varied the values of all 
the parameters and considered a gamma distribution with mean $3$ and variance 
$1.5$ for choosing values for $t_{ij}$. Also, we considered an extended model 
in which interactions between $t_{ij}$ and the dummy variables were included. 
In this case, we tested the interactions effects. For the sake of brevity, 
the results are not shown. In short, the Cox--Reid and the two Bartlett-corrected 
tests outperformed $LR$. For instance, our simulation experiment with  
$\beta_0 = 0.2$, $\beta_1 = 0.4$, $\beta_2 = \beta_3 = 0$, $\omega_1 = 1.5$,
$\omega_2 = 0.05$, $\omega_3 = 1.2$, $\omega_4 = 0.10$ and $N = 24$ yielded 
the following null rejection rates at the 10\% nominal level:
$14.7\%$ $(LR)$, $10.6\%$ $(LR^*)$, $9.4\%$ $(LR_{CR})$ and
$9.8\%$ $(LR^*_{CR})$. Also, for an extended model which includes
two parameters, $\beta_4$ and $\beta_5$, representing interactions
between $t_{ij}$ and the dummy variables, we obtained
$7.1\%$ $(LR)$, $5.4\%$ $(LR^*)$, $3.3\%$ $(LR_{CR})$ and $5.0\%$ 
$(LR^*_{CR})$ for $\alpha = 5\%$ and $N = 24$.
Here, $\beta_0 = 0.2$, $\beta_1 = 0.4$, $\beta_2 = 0.3$, $\beta_3 = 0.5$, 
$\beta_4 = \beta_5 = 0$ and the same values for $\omega_1, \ldots, \omega_4$ as before.

\section{Blood pressure data}\label{S:application}

We shall now present an application that uses a real data set. 
The data consist of a 
randomly selected subset of the data used by Crepeau et al.~(1985).  
Heart attacks were induced in rats exposed to four different low concentrations of 
halothane; group 1: 0\% (control), group 2: 0.25\%, group 3: 0.50\% and group 4: 1.0\%. 
Our sample consists of 23 rats. The blood pressure of each rat (in mm Hg) is 
recorded over different points in time, from 1 to 9 recordings, after the induced 
heart attack. The main goal is to investigate the effect of halothane on the blood 
pressure.

Figure 2 shows plots of blood pressure versus time for each rat. Clearly, the
profiles differ on the intercept. However, the slopes are not markedly different.
At the outset, we consider a model where blood pressure varies linearly with
time, possibly with different intercepts and slopes for each concentration of
halothane, and with intercept and slope random effects to account for
animal-to-animal variation. As we will see later, the usual likelihood
ratio test rejects the null hypothesis of
common slope at the 10\% nominal level, unlike the modified tests.

\begin{figure}[h]
\centering  
\includegraphics[width=10cm, height=8cm]{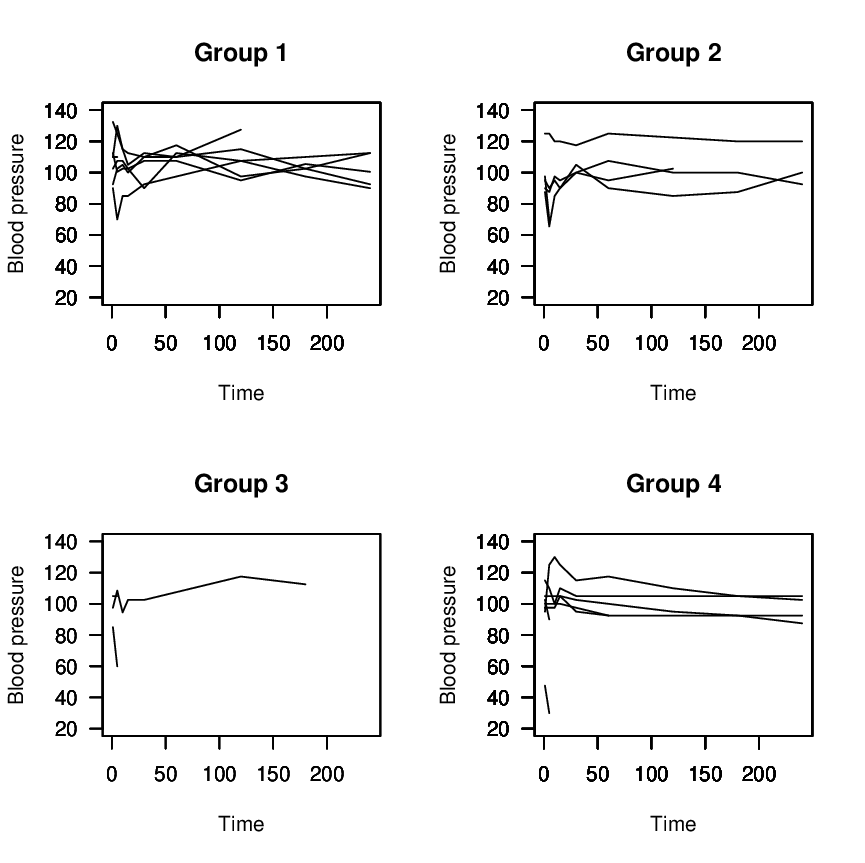}
\caption{\footnotesize Blood pressure against time for each rat.}\label{fig2}
\end{figure}

The mixed linear model considered here is 
\begin{equation}%\label{ModReal1}
\begin{split}
y_{ij} &= \beta_0 + \beta_1 t_{ij} + \gamma_{02} G_{2i} + \gamma_{03} G_{3i} + \gamma_{04} G_{4i}
+ \gamma_{12} G_{2i} t_{ij} + \gamma_{13} G_{3i} t_{ij} \\
&+ \gamma_{14} G_{4i} t_{ij} + b_{0i} + b_{1i} t_{ij} + \epsilon_{ij},
\end{split}
\end{equation}
with $i = 1, 2, \ldots, 23$ and $j = 1, 2, \ldots, \tau_i$, where $y_{ij}$ is 
the blood pressure of the $i$th rat at time $j$, 
$t_{ij}$ is the $j$th point in time (in minutes) in which the $i$th rat blood 
pressure was recorded, and $G_{2i}$ is a dummy variable that equals 1 if 
the $i$th rat belongs to group 2 and 0 otherwise. Also, 
$G_{3i}$ and $G_{4i}$ equal 1 for groups 3 and 4, respectively. 
We assume that $b_i = (b_{0i} \:\: b_{1i})^\top \, \stackrel{\mathrm{i.i.d.}}{\sim}\,
\mathcal{N}_2({\bf 0}, G)$, where $G$ is given in (\ref{MatrizG}).
Additionally, $\epsilon_{ij}\,\stackrel{\mathrm{i.i.d.}}{\sim}\,\mathcal{N}(0,\omega_4)$,
the $\epsilon_{ij}$'s being independent of the $b_i$'s. 

The maximum likelihood estimates of the fixed effects parameters are 
$\widehat\beta_0 = 104.360$, $\widehat\beta_1 = 0.004$, $\widehat\gamma_{02} = -0.719$, 
$\widehat\gamma_{03} = 0.203$, $\widehat\gamma_{04} = -15.211$,
$\widehat\gamma_{12} = 0.022$, $\widehat\gamma_{13} = 0.109$ 
and $\widehat\gamma_{14} = -0.019$. We wish to make inference on $\gamma_{12}$, 
$\gamma_{13}$ and $\gamma_{14}$. More specifically, we wish to test 
${\mathcal H}_0: \psi = {\bf 0}$ against ${\mathcal H}_1: \psi \neq {\bf 0}$, where
$\psi = (\gamma_{12}, \gamma_{13}, \gamma_{14})^\top$. Note that under 
the null hypothesis, the mean slopes are equal for the different halothane
concentrations. The adjusted profile maximum likelihood estimates of
$\gamma_{12} $, $\gamma_{13}$ and $\gamma_{14}$ are $\widetilde\gamma_{12} = 0.020$,
$\widetilde\gamma_{13} = 0.101$ and $\widetilde\gamma_{14} = -0.030$, respectively. 
The test statistics assume the following values: 
$LR = 6.522$ ($p$-value: 0.089), $LR^* = 5.678$ ($p$-value: 0.128), $LR_{a} = 5.287$ 
($p$-value: $0.152$) and ${LR^*_{CR}} = 6.168$ ($p$-value: $0.104$). 
The standard likelihood ratio test rejects the null hypothesis at the 10\% 
nominal level, i.e., it suggests that there are differences in mean slopes for
different dosages. The three modified tests,
however, suggest otherwise, i.e., the null 
hypothesis is not rejected by these tests at the same nominal level. 

We now consider the following reduced model:  
\begin{equation*}\label{ModReal2}
y_{ij} = \beta_0 + \beta_1 t_{ij} + \gamma_{02} G_{2i} + \gamma_{03} G_{3i} + \gamma_{04} G_{4i} + b_{0i}
+ b_{1i} t_{ij} + \epsilon_{ij},
\end{equation*}
with $i = 1, 2, \ldots, 23$ and $j = 1, 2, \ldots, \tau_i$. 
We wish to test 
${\mathcal H}^*_0: \psi^* = {\bf 0}$ against ${\mathcal H}^*_1: \psi^* \neq {\bf 0}$, 
where $\psi^* = (\gamma_{02}, \gamma_{03}, \gamma_{04})^\top$.
Note that we are testing whether the mean blood pressures are equal
across the different dosages. 
The fixed effects maximum likelihood estimates are 
$\widehat\beta_0 = 99.531$, $\widehat\beta_1 = 0.006$,
$\widehat\gamma_{02} = -0.525$, $\widehat\gamma_{03} = 2.318$ 
and $\widehat\gamma_{04} = -13.357$. The adjusted profile maximum likelihood 
estimates of $\gamma_{02} $, $\gamma_{03}$ and $\gamma_{04}$ are, 
respectively, $\widetilde\gamma_{02} = -0.823$, $\widetilde\gamma_{03} = 2.079$ 
and $\widetilde\gamma_{04} = -12.573$. We now have 
$LR = 6.143$ ($p$-value: $0.105$), $LR^* = 5.174$ ($p$-value: $0.159$), $LR_{a} = 4.002$
($p$-value: $0.261$) and ${LR^*_{CR}} = 4.167$ ($p$-value: $0.244$). All tests 
yield the same inference, namely: the null hypothesis is not rejected at the 
10\% nominal level.

Therefore, we conclude that there is no group effect.
In other words, the analysis carried out using the modified tests suggests that
the blood pressure is not affected by the administration of halothane at the
concentrations considered in the experiment.
This conclusion agrees with the findings of Crepeau et al.~(1985).

\section{Concluding remarks}\label{S:conclusions} 

We addressed the issue of performing likelihood-based testing inference on the 
fixed effects parameters of mixed linear models when the sample contains a small 
number of observations. 
The standard likelihood ratio test is liberal, as evidenced by 
our Monte Carlo results. We obtained three alternative tests, namely: an 
adjusted profile likelihood ratio test, its Bartlett-corrected version and 
also the Bartlett-corrected likelihood ratio test. Our results generalize those 
in Zucker et al.~(2000) in two directions. First,
we allow practitioners to test joint restrictions 
on one or more fixed effects parameters, whereas their results only hold for 
tests on a parameter at a time. Second, unlike Zucker et al.~(2000), we
do not assume that the covariance matrix of the random effects is linear
when deriving the Bartlett correction to the profile likelihood ratio test.
Our main results are stated through closed-form formulas that only involve
simple operations on vectors 
and matrices, and hence they can be easily implemented in matrix programming 
languages and statistical software. 
The simulation study we report clearly 
show that the proposed tests outperform the standard likelihood ratio test, 
especially when the sample size is small. It shows that the three alternative 
tests yield reliable inferences even for unbalanced data. In particular, 
the adjusted profile likelihood ratio test and its Bartlett-corrected version 
improve the type I error rate, especially when the number of 
observations is small.

\section*{Acknowledgements}

We gratefully acknowledge financial support from FAPESP and CNPq.
We also thank three anonymous referees for helpful suggestions.

\appendix
{\small
\section*{Appendix A. Derivation of $C$}\label{S:appendixA} 
We use the following tensor notation for log-likelihood cumulants:  
\begin{equation*}
\kappa_{rs} = {\rm E}\left(\frac{\partial^2 {\ell}}{\partial{\bf\vartheta}_r 
\partial{\bf\vartheta}_s}\right){\rm ,}\:\:\:
\:\kappa_{rst} = {\rm E}\left(\frac{\partial^3 {\ell}}{\partial{\bf\vartheta}_r 
\partial{\bf\vartheta}_s \partial{\bf\vartheta}_t}\right)\:\:\:{\rm and}\:\:\:\:\:
\kappa_{rstu} = {\rm E}\left(\frac{\partial^4 {\ell}}{\partial{\bf\vartheta}_r 
\partial{\bf\vartheta}_s \partial{\bf\vartheta}_t \partial{\bf\vartheta}_u}\right),
\end{equation*}
${\bf\vartheta}_r$ being the $r$th element of $\vartheta$. The notation used 
for derivatives of cumulants is the following: 
$$
\left(\kappa_{rs}\right)_{t} = \frac{\partial\kappa_{rs}}{\partial{\bf\vartheta}_t}, 
\:\:\: \left(\kappa_{rst}\right)_{u} = \frac{\partial\kappa_{rst}}{\partial{\bf\vartheta}_u}
\:\:\:{\rm and}\:\:\: \left(\kappa_{rs}\right)_{tu} = \frac{\partial\kappa_{rs}}
{\partial{\bf\vartheta}_t \partial{\bf\vartheta}_u}. 
$$
In what follows, we shall use similar notation for derivatives of matrices formed out 
of cumulants. Note that $-\kappa_{rs}$ is the $(r,s)$ element of Fisher's information matrix;
the $(r,s)$ element of its inverse is denoted by $-\kappa^{rs}$.

Lawley's~(1956) formula for $C$ is 
\begin{equation*}
C = \sum_{\psi, \xi, \omega} (l_{rstu} - l_{rstuvw})-\sum_{\xi, \omega} (l_{rstu} - 
l_{rstuvw}) = C_1 - C_2,
\end{equation*}
where $C_1 = \sum_{\psi, \xi, \omega} l_{rstu} - \sum_{\xi, \omega} l_{rstu}$ and
$C_2 = \sum_{\psi, \xi, \omega} l_{rstuvw}  - \sum_{\xi, \omega} l_{rstuvw}$ with
\begin{equation*}
l_{rstu} = \kappa^{rs} \kappa^{tu} \left\{\frac{1}{4} \kappa_{rstu} - 
\left(\kappa_{rst}\right)_{u} -  \left(\kappa_{rt}\right)_{su}\right\}
\nonumber
\end{equation*}
and
\begin{equation*}
\begin{split}
l_{rstuvw} &= \kappa^{rs} \kappa^{tu} \kappa^{vw} \Bigg\{\kappa_{rtv} 
\left(\frac{1}{6} \kappa_{suw} - \left(\kappa_{sw}\right)_{u}\right) + 
\kappa_{rtu} \left(\frac{1}{4} \kappa_{svw} - \left(\kappa_{sw}\right)_{v}\right)\\
&+ \left(\kappa_{rt}\right)_{v}\left(\kappa_{sw}\right)_{u} + \left(\kappa_{rt}\right)_{u}
\left(\kappa_{sw}\right)_{v} \Bigg\},
\end{split}
\end{equation*}
where the indices $r, s, t, u, v, w$ refer to the components of 
$\vartheta =({\bf\psi}^\top, {\bf\xi}^\top, {\bf\omega}^\top)^\top$. 
Here, $\sum_{\psi, \xi, \omega}$ denotes summation over 
all possible combinations of the $n+m+1$ parameters in $\vartheta$,  
and $\sum_{\xi, \omega}$ denotes summation over the combinations of the 
$n-p+m+1$ parameters in $(\xi^\top, \omega^\top)^\top$. We use indices $a, b, c, d$ 
in reference to the components of $\psi$, indices $f, g$ for the components of $\xi$, and 
indices $j, k, l, o$ for the elements of $\omega$. Further notation used here is
given in Sections \ref{S:mixed} and \ref{S:corrections}.

The first-order derivatives of the log-likelihood function in (\ref{E.2.5.1}) are 
\begin{equation*}
\begin{split}
\frac{\partial {\ell}(\vartheta;{\bf Y})}{\partial{\bf\psi}} &= \widetilde{X}_p'^\top
{\bf \Sigma^{-1}}{\bf z}, \:\:\: \frac{\partial {\ell}(\vartheta;{\bf Y})}{\partial{\bf\xi}} 
= \widetilde{X}_{n-p}^\top{\bf \Sigma^{-1}}{\bf z},\\
\frac{\partial 
{\ell}(\vartheta;{\bf Y})}{\partial{\bf\omega}_j} &= -\frac{1}{2}\tr({\bf \Sigma^{-1}}
{\bf \dot{\Sigma}}_j) - \frac{1}{2} {\bf z}^\top{\bf \dot{\Sigma}}^j {\bf z} +
{\bf\psi}^\top \dot{{X}}_j'^\top {\bf \Sigma^{-1}}{\bf z}.
\end{split}
\end{equation*}
The second-order derivatives are 
\begin{equation*}
\begin{split}
\frac{\partial^2 {\ell}(\vartheta;{\bf Y})}{\partial{\bf\psi}\partial{\bf\psi}^\top} &= 
- \widetilde{X}_p'^\top {\bf \Sigma^{-1}} \widetilde{X}_p', \ 
\frac{\partial^2 {\ell}(\vartheta;{\bf Y})}{\partial{\bf\xi}\partial{\bf\xi}^\top} = 
- \widetilde{X}_{n-p}^\top {\bf \Sigma^{-1}} \widetilde{X}_{n-p}, \
\frac{\partial^2 {\ell}(\vartheta;{\bf Y})}{\partial{\bf\psi}\partial{\bf\xi}^\top} = 
{\bf 0},  \\
\frac{\partial^2 {\ell}(\vartheta;{\bf Y})}{\partial{\bf\psi}\partial{\bf\omega}_j} &= 
(\dot{{X}}_j'^\top {\bf \Sigma^{-1}} + \widetilde{X}_p'^\top
{\bf \dot{\Sigma}}^j){\bf z}, \ 
\frac{\partial^2 {\ell}(\vartheta;{\bf Y})} {\partial{\bf\xi}\partial{\bf\omega}_j} = 
\widetilde{X}_{n-p}^\top {\bf \dot{\Sigma}}^j ({\bf Y} - \widetilde{X}_{n-p}{\bf\xi}), \\ 
\frac{\partial^2 {\ell}(\vartheta;{\bf Y})}{\partial{\bf\omega}_j \partial{\bf\omega}_k} &= 
-\frac{1}{2}\:\tr({\bf \dot{\Sigma}}^j {\bf \dot{\Sigma}}_k) -\frac{1}{2}\:\tr({\bf \Sigma^{-1}}
{\bf \ddot{\Sigma}}_{jk}) - {\bf\psi}^\top {\dot{X}_k'^\top} {\bf \Sigma^{-1}} 
\dot{{X}_j'} {\bf\psi} + {\bf\psi}^\top(\ddot{X}_{jk}'^\top{\bf \Sigma^{-1}} \\
&+ {\dot{X}_k'^\top} {\bf \dot{\Sigma}}^j + {\dot{X}_j'^\top} {\bf \dot{\Sigma}}^k) {\bf z} - 
\frac{1}{2} {\bf z}^\top {\bf \ddot{\Sigma}}^{jk} {\bf z}, 
\end{split}
\end{equation*}
where 
\begin{equation*}
\begin{split}
\ddot{X}_{jk}' = \frac{\partial \dot{{X}_j'}}{\partial {\bf\omega}_k} 
              &= 2 \widetilde{X}_{n-p}(\widetilde{X}_{n-p}^\top{\bf \Sigma^{-1}}
                 \widetilde{X}_{n-p})^{-1}\widetilde{X}_{n-p}^\top {\bf \dot{\Sigma}}^k \widetilde{X}_{n-p}
                 (\widetilde{X}_{n-p}^\top{\bf \Sigma^{-1}}\widetilde{X}_{n-p})^{-1}\widetilde{X}_{n-p}^\top 
                 {\bf \dot{\Sigma}}^j \widetilde{X}_p' \\
&- \widetilde{X}_{n-p}(\widetilde{X}_{n-p}^\top{\bf \Sigma^{-1}}\widetilde{X}_{n-p})^{-1}
\widetilde{X}_{n-p}^\top {\bf \ddot{\Sigma}}^{jk} \widetilde{X}_p'.
\end{split}
\end{equation*} 
Additionally, the third-order derivatives are 
\begin{equation*}
\begin{split}
\frac{\partial^3 {\ell}(\vartheta;{\bf Y})}{\partial{\bf\xi}\partial{\bf\xi}^\top 
\partial{\bf\omega}_j} &= -\widetilde{X}_{n-p}^\top {\bf \dot{\Sigma}}^j \widetilde{X}_{n-p}, 
\ \
\frac{\partial^3 {\ell}(\vartheta;{\bf Y})}{\partial{\bf\psi}\partial{\bf\psi}^\top 
\partial{\bf\omega}_j} 
= - \widetilde{X}_p'^\top {\bf \dot{\Sigma}}^j \widetilde{X}_p',
\\
\frac{\partial^3 {\ell}(\vartheta;{\bf Y})}{\partial{\bf\psi}\partial{\bf\xi}^\top 
\partial{\bf\xi}_f} 
&= \frac{\partial^3 {\ell}(\vartheta;{\bf Y})}{\partial{\bf\psi}\partial{\bf\xi}^\top 
\partial{\bf\omega}_j} = {\bf 0},
\ \ 
\frac{\partial^3 {\ell}(\vartheta;{\bf Y})}{\partial{\bf\xi} \partial{\bf\omega}_j 
\partial{\bf\omega}_k}
= \widetilde{X}_{n-p}^\top {\bf \ddot{\Sigma}}^{jk} ({\bf Y} - \widetilde{X}_{n-p}{\bf\xi}),
\\ 
\frac{\partial^3 {\ell}(\vartheta;{\bf Y})}{\partial{\bf\psi}\partial{\bf\omega}_j 
\partial{\bf\omega}_k} 
&= (\ddot{X}_{jk}'^\top{\bf \Sigma^{-1}} + {\dot{X}_k'^\top} 
{\bf \dot{\Sigma}}^j + {\dot{X}_j'^\top} {\bf \dot{\Sigma}}^k 
+ \widetilde{X}_p'^\top {\bf \ddot{\Sigma}}^{jk}) {\bf z}, 
\\
\frac{\partial^3 {\ell}(\vartheta;{\bf Y})}{\partial{\bf\omega}_j \partial{\bf\omega}_k 
\partial{\bf\omega}_l} 
= &-\frac{1}{2}(\:\tr({\bf \ddot{\Sigma}}^{lk} {\bf \dot{\Sigma}}_j) + \tr({\bf 
\dot{\Sigma}}^k {\bf \ddot{\Sigma}}_{lj}) + \tr({\bf \dot{\Sigma}}^l {\bf 
\ddot{\Sigma}}_{jk}) + \tr({\bf \Sigma^{-1}} {\bf \ddot{\Sigma}}_{jkl})
+{\bf z}^\top {\bf \ddot{\Sigma}}_{jkl} {\bf z}) 
\\
&+ {\bf \psi}^\top (\ddot{X}_{lk}'^\top {\bf \dot{\Sigma}}^{j}  
+ {\dot{X}_k'^\top} {\bf \ddot{\Sigma}}^{lj} + \ddot{X}_{jk}'^\top 
{\bf \dot{\Sigma}}^{l} + \ddot{X}_{lj}'^\top {\bf \dot{\Sigma}}^{k} 
+ {\dot{X}_j'^\top} {\bf \ddot{\Sigma}}^{lk} + {\dot{X}_l'^\top} 
{\bf \ddot{\Sigma}}^{kj} \\
&+ \ddot{X}_{jkl}'^\top {\bf \Sigma}^{-1}) {\bf z}, 
\end{split}
\end{equation*}  
where ${\bf \ddot{\Sigma}}_{jkl} = \partial{\bf \ddot{\Sigma}}_{jk}/\partial{\bf\omega}_l$ 
and $\ddot{X}_{jkl}' = \partial\ddot{X}_{jk}'/\partial{\bf\omega}_l$. Finally, 
the fourth-order derivatives can be shown to be 
\begin{equation*}
\begin{split}
\frac{\partial^4 {\ell}(\vartheta;{\bf Y})}{\partial{\bf\psi}\partial{\bf\psi}^\top 
\partial{\bf\omega}_j \partial{\bf\omega}_k} &= -2 {\dot{X}_k'^\top}
{\bf \dot{\Sigma}}^{j} \widetilde{X}_p' - \widetilde{X}_p'^\top {\bf \ddot{\Sigma}}^{jk} 
\widetilde{X}_p',\\
\frac{\partial^4 {\ell}(\vartheta;{\bf Y})}{\partial{\bf\psi}\partial{\bf\psi}^\top 
\partial{\bf\xi}_f \partial{\bf\xi}_g} &= \frac{\partial^4 {\ell}(\vartheta;{\bf Y})}{\partial{\bf\psi}\partial{\bf\psi}^\top \partial{\bf\xi}_f \partial{\bf\omega}_j} = 
\frac{\partial^3 {\ell}(\vartheta;{\bf Y})}{\partial{\bf\psi}\partial{\bf\psi}^\top 
\partial{\bf\xi}_f} = {\bf 0}.
\end{split}
\end{equation*}   
Taking expected values of second, third and fourth derivatives, we obtain 
\begin{equation*}
\begin{split}
K_{{\bf \psi}{\bf \psi}} &=
     {\rm E}\left(\frac{\partial^2 {\ell}(\vartheta;{\bf Y})}
                       {\partial{\bf\psi}\partial{\bf\psi}^\top} \right) =
     -\widetilde{X}_p'^\top 
     {\bf \Sigma^{-1}} \widetilde{X}_p', \\
K_{{{\bf \xi}{\bf \xi}}{\bf \omega}_j} &=
     {\rm E}\left(\frac{\partial^3 {\ell}(\vartheta;{\bf Y})}
                       {\partial{\bf\xi} \partial{\bf\xi}^\top \partial{\bf \omega}_j}\right) =
     - \widetilde{X}_{n-p}^\top {\bf\dot{\Sigma}}^{j} \widetilde{X}_{n-p},\\
K_{{\bf \psi}{\bf \psi}{\bf \omega}_j {\bf \omega}_k} &=
     {\rm E}\left(\frac{\partial^4 {\ell}(\vartheta;{\bf Y})}
                  {\partial{\bf\psi}\partial{\bf\psi}^\top \partial{\bf \omega}_j
                  \partial{\bf \omega}_k}\right) =
     -2 {\dot{X}_k'^\top} {\bf \dot{\Sigma}}^{j} \widetilde{X}_p'
     - \widetilde{X}_p'^\top {\bf \ddot{\Sigma}}^{jk} \widetilde{X}_p'.
\end{split}
\end{equation*}  
In similar fashion, it follows that 
\begin{equation*}
\begin{split}
K_{{\bf \xi}{\bf \xi}} &= -\widetilde{X}_{n-p}^\top {\bf \Sigma^{-1}} \widetilde{X}_{n-p},\ \
K_{{\bf \xi}{\bf \omega}_j} = \widetilde{X}_{n-p}^\top {\bf \dot{\Sigma}}^{j} 
\widetilde{X}_p'\: {\bf \psi}, \ \
K_{{\bf \psi}{\bf \omega}_j} = {\bf 0},\\
K_{{\bf \psi}{\bf \psi}{\bf \omega}_j} &= - \widetilde{X}_p'^\top {\bf\dot{\Sigma}}^{j} 
\widetilde{X}_p', \ \
K_{{\bf \psi}{\bf \xi}{\bf \xi}_f} = K_{{\bf \psi}{\bf \xi}{\bf \omega}_j} = {\bf 0},\\
K_{{\bf \psi}{\bf \omega}_j{\bf \omega}_k} &= {\bf 0},\ \
K_{{\bf \xi} {\bf \omega}_j {\bf \omega}_k} = \widetilde{X}_{n-p}^\top  {\bf \ddot{\Sigma}}^{jk} 
\widetilde{X}_p'\: {\bf \psi}, \ \
K_{{\bf \psi}{\bf \psi}{\bf \xi}_f {\bf \xi}_g} = K_{{\bf \psi} {\bf \psi}{\bf \xi}_f 
{\bf \omega}_j} = {\bf 0}.
\end{split}
\end{equation*}  
Additionally, 
\begin{equation*}
\begin{split}
\kappa_{jk} &= \frac{1}{2}\tr({\bf \dot{\Sigma}}^{j} {\bf \dot{\Sigma}}_{k}) - 
{\bf \psi}^\top {\dot{X}_j'^\top} {\bf \Sigma^{-1}} \dot{{X}_{k}'} 
{\bf \psi}, \\
\kappa_{ljk} &= -2\:\tr({\bf \dot{\Sigma}}^{l} {\bf \dot{\Sigma}}_{k} 
{\bf \Sigma^{-1}}{\bf \dot{\Sigma}}_{j}) + \frac{1}{2}\:\tr({\bf \dot{\Sigma}}^j 
{\bf \ddot{\Sigma}}_{lk})+ \frac{1}{2}\: \tr({\bf \dot{\Sigma}}^k {\bf \ddot{\Sigma}}_{lj}) 
+ \frac{1}{2}\:\tr({\bf \dot{\Sigma}}^l {\bf \ddot{\Sigma}}_{jk}). 
\end{split}
\end{equation*}   
Consider the following matrices formed out of minus Fisher's information inverse:  
$K^{\psi\psi} = K_{\psi\psi}^{-1}$, $K^{\omega \omega} = (K_{\omega \omega} - 
K_{\xi\omega}^\top K_{\xi\xi}^{-1} K_{\xi\omega})^{-1}$, $K^{\xi\xi} = K_{\xi\xi}^{-1} +  
K_{\xi\xi}^{-1} K_{\xi\omega} K^{\omega \omega} K_{\xi\omega}^\top K_{\xi\xi}^{-1}$ and
$K^{\xi\omega} = {K^{\omega\xi}}^\top = -K_{\xi\xi}^{-1} K_{\xi\omega} {K^{\omega\omega}}^\top$,
where the $j$th column of $K_{\xi\omega}$ is $K_{{\bf \xi}{\bf \omega}_j}$ and the
$(j,k)$th element of $K_{\omega\omega}$ is $\kappa_{jk}$. 
It can be shown that 
\begin{equation*}
\begin{split}
\left(K_{{\bf \psi}{\bf \psi}}\right)_{j} &= - \widetilde{X}_p'^\top {\bf\dot
{\Sigma}}^{j} \widetilde{X}_p', \ \
\left(K_{{\bf \psi}{\bf \psi}}\right)_{jk} = -2 {\dot{X}_k'^\top} 
{\bf \dot{\Sigma}}^j {\widetilde{X}_{p}'} - \widetilde{X}_{p}'^\top {\bf \ddot{\Sigma}}^{jk} 
{\widetilde{X}_{p}'},
\\
\left(K_{{\bf \xi}{\bf \xi}}\right)_{j} &= -\widetilde{X}_{n-p}^\top {\bf\dot{\Sigma}}^{j} 
\widetilde{X}_{n-p},\ \
(K_{{\bf \xi}{\bf \omega}_j})_{k} = \widetilde{X}_{n-p}^\top {\bf \ddot{\Sigma}}^{jk} 
\widetilde{X}_p'\: {\bf \psi} + \widetilde{X}_{n-p}^\top {\bf \dot{\Sigma}}^{j} 
\dot{{X}_{k}'}\: {\bf \psi}, \\
(\kappa_{jl})_{k} &= - \tr({\bf \dot{\Sigma}}^{l} {\bf \dot{\Sigma}}_{j} 
{\bf \Sigma^{-1}} {\bf \dot{\Sigma}}_{k}) + \frac{1}{2} \tr({\bf \dot{\Sigma}}^{j} 
{\bf \ddot{\Sigma}}_{lk}) + \frac{1}{2}\tr({\bf \dot{\Sigma}}^{l} {\bf \ddot{\Sigma}}_{jk}).
\end{split}
\end{equation*}  
It follows from the orthogonality between $\psi$ and $(\xi^\top,\omega^\top)^\top$ 
that $\kappa^{af} = \kappa^{aj} = (\kappa_{af})_{jb} = (\kappa_{aj})_{fb} = 0$. 
Also, $\kappa_{jfa} =  \kappa_{jfab} = 0$. Hence, 
\begin{equation*}
\begin{split}
C_1 = \sum\left( l_{abcd} + l_{abfg} + l_{abfj} + l_{abjf} + l_{abjk} + l_{fgab} 
+ l_{jkab}\right),
\end{split}
\end{equation*}  
where $\sum$ ranges over all parameter combinations induced by the indices 
$a,b,c,d,f,g,\break j,k$. It is possible to show that $l_{abcd} = l_{abfg} 
= l_{abfj} = l_{abjf} = l_{fgab} = 0$. Thus, 
\begin{equation*}
\begin{split}
C_1 = \sum\left(l_{abjk} +  l_{jkab}\right) = \sum\left\{\kappa^{ab} \kappa^{jk} 
\left(\frac{1}{4}\kappa_{abjk} - \left(\kappa_{abj}\right)_{k} \right) + 
\frac{1}{4}\kappa^{jk} \kappa^{ab} \kappa_{jkab} \right\}.
\end{split}
\end{equation*}
Since $\kappa_{abjk} = \left(\kappa_{abj}\right)_{k} = \kappa_{jkab}$, $C_1$ reduces to 
\begin{equation*}
\begin{split}
C_1 = -\frac{1}{2} \sum\kappa^{ab} \kappa^{jk} \kappa_{abjk}.
\end{split}
\end{equation*}

As for $C_2$, we have that 
\begin{equation*}
\begin{split}
C_2 &= \sum(l_{abcdjk} + l_{abjkcd} + l_{jkloab} + l_{jkablo} 
+ l_{jkabcd} + l_{fjgkab} + l_{fjkgab} + l_{fjklab} \\
&+ l_{jkfgab} + l_{jkflab} + l_{jklfab}  + l_{jfabgk} 
+ l_{jfabkg}+ l_{jfabkl}+ l_{jkabfg} + l_{jkabfl} + l_{jkablf}) 
\\ \\ 
&= \sum\Big\{-\frac{1}{4} \kappa^{ab} \kappa^{cd} \kappa^{jk}  \kappa_{abj} 
\kappa_{cdk} + \frac{1}{2} \kappa^{ab} \kappa^{fg} \kappa^{jk} \kappa_{abj} 
\kappa_{fgk} \\ 
&\ \  \ \ \ \ \ \ \ \ \ - \kappa^{ab} \kappa^{fj} \kappa^{kl} \kappa_{abj}\left(2 
(\kappa_{fk})_{l} - \frac{3}{2} \kappa_{fkl} \right)
+ \frac{1}{2} \kappa^{ab} \kappa^{jk} \kappa^{lo} \kappa_{abj}\big(\kappa_{klo} 
- 2(\kappa_{kl})_{o} \big)\Big\}. 
\end{split}
\end{equation*}  
Therefore, $C$ reduces to 
\begin{equation*}
\begin{split}
 C &= \sum \Big\{-\frac{1}{2} \kappa^{ab} \kappa^{jk} \kappa_{a b j k} + 
 \frac{1}{4} \kappa^{ab} \kappa^{cd} \kappa^{jk} \kappa_{abj} \kappa_{cdk} 
- \frac{1}{2}  \kappa^{ab} \kappa^{jk} \kappa^{lo} \kappa_{abj}
\left(\kappa_{lok} - 2(\kappa_{lo})_{k} \right)
\\ 
&+ \kappa^{ab} \kappa^{fk} \kappa^{jl} \kappa_{abj}\Big(2 (\kappa_{fk})_{l} 
- \frac{3}{2} \kappa_{fkl} \Big) - \frac{1}{2} \kappa^{ab} \kappa^{fg} 
\kappa^{jk} \kappa_{abj} \kappa_{fgk}\Big\}. 
\end{split}
\end{equation*}  
We now arrive at the matrix expression given by 
\begin{equation*}\label{E.A.1}
C = \tr\left( K^{\omega\omega}\left\{-\frac{1}{2} M + \frac{1}{4} P
- \left(  \frac{1}{2}\:\rho - \delta + \frac{1}{2}\:\eta\right)\tau^\top\right\}\right).
\eqno{(\rm{A}.1)}
\end{equation*}

Here, $\rho$, $\delta$ and 
$\eta$ are $(m+1)$-vectors whose $j$th elements are, respectively, 
$\tr(K^{\omega\omega} A^{(j)})$, $\tr({K^{\xi\omega}}^\top B^{(j)})$ and
$\tr(-K^{\xi\xi}(\widetilde{X}_{n-p}^\top {\bf \dot{\Sigma}}^{j} \widetilde{X}_{n-p}))$. 
In our notation, 
$B^{(j)}$ is a matrix that contains the $m+1$ column vectors  
$(1/2\:\widetilde{X}_{n-p}^\top {\bf \ddot{\Sigma}}^{jk} 
\widetilde{X}_p' + 2\:\widetilde{X}_{n-p}^\top {\bf \dot{\Sigma}}^{j}\dot{{X}_{k}'})\psi$
and $A^{(j)}$ is defined in Section \ref{S:corrections1}. 
For testing ${\mathcal H}_0: \psi= {\bf 0}$, $C$ reduces to equation
$(\ref{Eq.9})$.

\section*{Appendix B. Derivation of $C^*$}\label{S:appendixB}

We shall now obtain $C^*$, which is used to Bartlett-correct the adjusted profile 
likelihood ratio test statistic. 
DiCiccio and Stern (1994, eq.\ (25)) give the following general expression:
\begin{equation*}\label{A1.3.1}
\begin{split}
C^* &= \sum_{\psi, \xi, \omega} \Bigg\{\frac{1}{4} \tau^{ru} \tau^{st} 
\kappa_{rstu} - \kappa^{ru} \tau^{st} (\kappa_{rst})_u + \Big(\kappa^{ru} 
\kappa^{st} - \nu\:^{ru} \nu\:^{st}\Big)(\kappa_{rs})_{tu} 
\\
&- \Big(\frac{1}{4} 
\kappa^{ru} \tau^{st} \tau^{vw} + \frac{1}{2} \kappa^{ru} \tau^{sw} \tau^{tv} 
-\frac{1}{3} \tau^{ru} \tau^{sw} \tau^{tv} \Big)\kappa_{rst} \kappa_{uvw} 
+ \Big(\kappa^{ru} \tau^{st} \kappa^{vw} + \kappa^{ru} \kappa^{sw} \kappa^{tv} 
\\
&- \nu\:^{ru} \kappa^{sw} \nu\:^{tv}\Big) \kappa_{rst} \left(\kappa_{uv}\right)_{w} 
- \Big(\kappa^{ru} \kappa^{st} \kappa^{vw} - \nu\:^{ru} \nu\:^{st} \nu\:^{vw}\Big) 
(\kappa_{rs})_{t} (\kappa_{uv})_{w} 
\\
&- \Big(\kappa^{ru} \kappa^{sw} \kappa^{tv} 
- \nu\:^{ru} \nu\:^{sw}\nu\:^{tv}\Big) 
(\kappa_{rs})_{t} (\kappa_{uv})_{w}\Bigg\}, 
\end{split}
\end{equation*}
where $\nu\:^{rs} = \kappa^{rs} - \tau^{rs}$, $\tau^{rs} = \kappa^{r b} 
\kappa^{s a} \sigma_{a b}$, $\sigma_{a b}$ being the $(a,b)$ element of the 
inverse of $K^{\psi \psi}$. From the orthogonality between $\psi$ and $\phi$ 
we have that $\tau^{fg} = \tau^{jk} = \tau^{fj} = \tau^{af} = \tau^{aj} = 0$. 
Also, $\tau^{ab} = \kappa^{ab}$. Thus, 
\begin{equation*}\label{A1.3.2}
\begin{split}
C^* &= \sum\Big\{\frac{1}{4}\kappa^{ad} \kappa^{bc} \kappa_{abcd} - \kappa^{ru} 
\kappa^{ab} (\kappa_{rab})_u + \left(\kappa^{ru} \kappa^{st} - \nu\:^{ru} \nu\:^{st}
\right)(\kappa_{rs})_{tu} \\
&- \Big(\frac{1}{4} \kappa^{ru} \kappa^{ab} \kappa^{cd} + 
\frac{1}{2} \kappa^{ru} \kappa^{ad} \kappa^{bc} \Big)\kappa_{rab} \kappa_{ucd}
+ \kappa^{ru} \kappa^{ab} \kappa^{vw}\kappa_{rab} \left(\kappa_{uv}\right)_{w} \\ 
&
+ \left(\kappa^{ru}\kappa^{tv} - \nu\:^{ru} \nu\:^{tv}\right) \kappa^{sw} \kappa_{rst} \left(\kappa_{uv}\right)_{w}\Big\}. 
\end{split}
\end{equation*}
We have that $\kappa^{ru}\kappa^{tv} - \nu\:^{ru} \nu\:^{tv} = \kappa^{ru} \tau^{tv} 
+ \kappa^{tv} \tau^{ru} - \tau^{ru} \tau^{tv}$
and $\left(\kappa_{bd}\right)_{k} = \kappa_{bdk}$. Hence,
$$\sum\left(\kappa^{ru}\kappa^{tv} 
- \nu\:^{ru} \nu\:^{tv}\right)\kappa^{sw} \kappa_{rst} \left(\kappa_{uv}\right)_{w} 
= \sum \kappa^{ab} \kappa^{cd} \kappa^{jk} \kappa_{ajc}\kappa_{bdk}.$$ 
Since $\kappa_{abcd} = \kappa_{abc} = \kappa_{fab} = (\kappa_{ac})_{bu} = (\kappa_{af})_{tu} 
=( \kappa_{aj})_{tu} = 0$,
it follows that $C^*$ reduces to  
\begin{equation*}\label{A1.3.6}
\begin{split}
C^* &= \sum\Big\{- \kappa^{ab} \kappa^{jk} \kappa_{abjk} + \frac{1}{4} \kappa^{ab} 
\kappa^{cd} \kappa^{jk} \kappa_{abj} \kappa_{cdk} + \kappa^{ab} \kappa^{jk} \kappa^{lo} 
\kappa_{abj} (\kappa_{kl})_{o} \\
&\ \ \ \ \ \ \ \ \ \ \ + \kappa^{ab} \kappa^{fj} \kappa^{gk} \kappa_{abj} \kappa_{fgk} 
+ 2 \kappa^{ab} \kappa^{fj} \kappa^{kl} \kappa_{abj} (\kappa_{fk})_{l}\Big\}. 
\end{split}
\end{equation*}
We then arrive at the matrix expression 
\begin{equation*}
C^* = \tr\left(K^{\omega\omega}\left\{ -M + \frac{1}{4} P
+ (\rho^* + 2\:\delta^*) \tau^\top\right\}\right)
+ \tau^\top K^{\omega\xi} \:\eta^*,
\eqno{(\rm{B}.1)}
\end{equation*}
where the $j$th elements of the vectors $\rho^*$ and $\delta^*$ 
are, respectively, $\tr(K^{\omega\omega} C^{(j)})$
and $\tr({K^{\xi\omega}}^\top F^{(j)})$, 
and the $f$th element of the vector $\eta^*$ is
$\tr(K^{\omega\xi} G^{(f)})$.
Also, $C^{(j)}$ is defined in Section \ref{S:corrections2}, 
$F^{(j)}$ is a matrix that contains the $m+1$ column vectors  
$\left(\widetilde{X}_{n-p}^\top {\bf \ddot{\Sigma}}^{jk} \widetilde{X}_p' +
\widetilde{X}_{n-p}^\top {\bf \dot{\Sigma}}^{j}\dot{{X}_{k}'}\right)\psi$
and $G^{(f)}$ is the $(n-p)\times(m+1)$ matrix whose $j$th column
is the $f$th column of $-\widetilde{X}_{n-p}^\top {\bf \dot{\Sigma}}^{j} \widetilde{X}_{n-p}$.
For testing ${\mathcal H}_0: \psi={\bf 0}$, $C^*$ reduces to equation
$(\ref{Eq.9.1})$.

}

\end{document}